\documentclass[12pt,a4paper,final]{iopart}

\usepackage{iopams}  
\usepackage{graphicx}
\usepackage{cite}
\usepackage[breaklinks=true,colorlinks=true,linkcolor=blue,urlcolor=blue,citecolor=blue]{hyperref}

\begin{document}

\title[]{Highly nonlinear ionization of atoms induced by intense high-harmonic pulses}

\author{B Senfftleben$^1$, M Kretschmar$^1$, A Hoffmann$^1$, M Sauppe$^{1,2}$, J T\"ummler$^1$, I Will$^1$, T Nagy$^1$, M J J Vrakking$^1$, D Rupp$^{1,2}$ and B Sch\"utte$^1$}

\address{$^1$Max-Born-Institut, Max-Born-Str. 2A, 12489 Berlin, Germany}
\address{$^2$Current address: ETH Zurich, John-von-Neumann-Weg 9, 8093 Zürich, Switzerland}
\ead{daniela.rupp@ethz.ch}
\ead{Bernd.Schuette@mbi-berlin.de}

\vspace{2pc}
\noindent{\it Keywords}: high-harmonic generation, extreme-ultraviolet pulses, nonlinear optics
%\submitto{\JPP}
%\maketitle

\section{Introduction}

\begin{abstract}
Intense extreme-ultraviolet (XUV) pulses enable the investigation of XUV-induced nonlinear processes and are a prerequisite for the development of attosecond pump - attosecond probe experiments. While highly nonlinear processes in the XUV range have been studied at free-electron lasers (FELs), high-harmonic generation (HHG) has allowed the investigation of low-order nonlinear processes. Here we suggest a concept to optimize the HHG intensity, which surprisingly requires a scaling of the experimental parameters that differs substantially from optimizing the HHG pulse energy. As a result, we are able to study highly nonlinear processes in the XUV range using a driving laser with a modest ($\approx 10$~mJ) pulse energy. We demonstrate our approach by ionizing Ar atoms up to Ar$^{5+}$, requiring the absorption of at least 10 XUV photons.
\end{abstract}

%\thispagestyle{fancy}
%\ifthenelse{\boolean{shortarticle}}{\abscontent}{}

%The generation and application of intense XUV pulses from FEL and HHG sources have enabled the experimental investigation of atoms, molecules, thin films, biological cells, clusters and solids with (sub-)femtosecond temporal resolution. 

Three key features of intense XUV pulses from FEL and HHG sources have opened up new possibilities for a range of fields from life science to material science and fundamental physics: (i) Intense XUV pulses provide the possibility to perform pump-probe experiments, in which a first XUV pump pulse initiates dynamics in an atom or a molecule, and these dynamics are probed by a second, time-delayed XUV probe pulse. This capability has been extensively used with femtosecond time resolution at FELs (see e.g. Ref.~\cite{jiang10}) and has been implemented with attosecond time resolution using HHG sources~\cite{tzallas11, takahashi13}. (ii) Intense, coherent XUV pulses have enabled single-shot coherent diffractive imaging (CDI) of isolated nanotargets with a resolution in the tens of nanometers range. While these experiments are predominantly performed at FELs~\cite{bogan08}, CDI on He nanodroplets using an intense HHG source was recently reported~\cite{rupp17}. (iii) High XUV intensities enable the study of nonlinear optics in this spectral range. Examples include multiple ionization of atoms~\cite{motomura09} and XUV-driven four-wave mixing schemes~\cite{foglia18}.

Advantages of HHG sources compared to FELs include their smaller size and lower costs resulting in easier access to these sources. Another important point is that two-color XUV-optical pump-probe experiments, which are one of the preferred applications of ultrashort XUV pulses, are challenging at FELs due to timing jitter between the optical and XUV beams~\cite{savelyev17}. Such experiments are routinely performed using HHG sources, with a time resolution extending into the attosecond domain, given the fact that attosecond pulses based on HHG have been generated for more than 10 years~\cite{krausz09}.

The most common scheme used to generate intense HHG pulses requires powerful near-infrared (NIR) pulses that are loosely focused into a gas medium to increase the generation volume~\cite{takahashi13,schutte14a,manschwetus16,nayak18,bergues18}. It has been shown that the HHG pulse energy increases with increasing NIR focal length~\cite{boutu11}. In contrast, we demonstrate in this paper that for a given laboratory size the XUV intensity on target can be optimized by using an NIR focusing element with a relatively short focal length. This enables the generation of a smaller XUV beam waist radius in the experiment and a higher XUV beamline transmission. As a result, we achieve an estimated XUV intensity in our laboratory of $7 \times 10^{14}$~W/cm$^2$, allowing us to study highly nonlinear processes, which we demonstrate in this paper by ionizing neutral Ar atoms up to Ar$^{5+}$.

\section{HHG intensity scaling}

Investigations on HHG scaling have shown that the HHG conversion efficiency may be preserved when varying the laser power, as long as all relevant parameters such as the focusing geometry and the gas pressure and length of the HHG medium are scaled appropriately~\cite{heyl16}. One can further assume that the NIR intensity used to optimize the HHG conversion efficiency is nearly independent of the chosen experimental conditions, given that the optimum NIR intensity is dictated by phase-matching conditions that are largely determined by the choice of laser wavelength and target gas. As a result, the XUV intensity obtained directly behind the generation medium is to a good approximation independent of the NIR focusing geometry. Therefore, in order to optimize the peak XUV intensity in the experiment, we apply the following strategy: (1) We choose a beamline geometry leading to a large demagnification $D=w_{XUV,source}/w_{XUV,focus}$ (where $w_{XUV,source}$ is the XUV beam waist radius at the HHG source and $w_{XUV,focus}$ is the XUV beam waist radius after focusing). (2) The HHG beamline transmission is optimized by using as few XUV optical elements (filters, mirrors) as possible, as each of these elements introduces high losses.

To study how the demagnification of the XUV beam radius from the HHG source to the XUV focus in the experiment depends on the NIR focal length $d_{NIR}$, we have performed calculations based on Gaussian beams in the paraxial approximation using ABCD matrices~\cite{siegman86}. While typical XUV beams from HHG may be non-Gaussian, it is known that certain fundamental correlations are equivalent for Gaussian and non-Gaussian beams~\cite{porras92}. In particular, the product of divergence and minimal beam diameter is preserved while propagating the beam through an ABCD optical system. Hence, using Gaussian optics is the most straightforward way to study scaling of the XUV intensity that can be achieved in an experiment.

\begin{figure}[tb]
	\centering
	\fbox{\includegraphics[width=\linewidth]{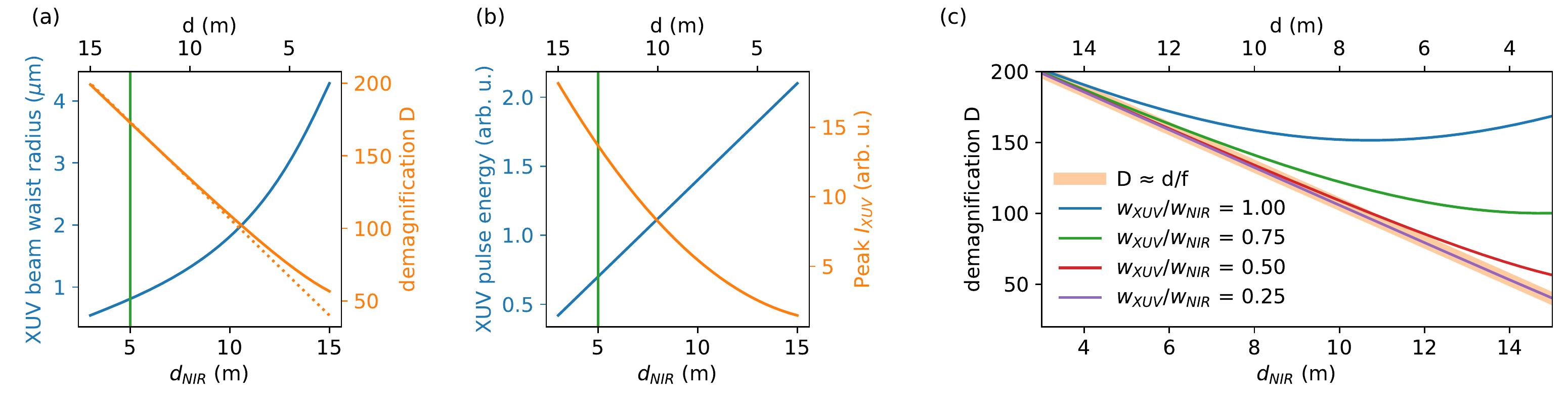}}
	\caption{Calculated scaling of the XUV parameters as a function of the NIR focal length $d_{NIR}$, assuming a total HHG beamline length of 18~m and an XUV focusing mirror with a focal length of 75~mm (corresponding to the conditions in our laboratory). (a) XUV beam waist radius (blue curve) and demagnification resulting from simulations using Gaussian beams (solid orange curve) and calculated according to $D\approx d/f_{XUV}$ (dotted orange curve), where $d$ is the distance from the HHG source to the XUV mirror. The green line corresponds to the position of the NIR focal plane in our setup. (b) Anticipated XUV pulse energy (blue curve) and XUV peak intensity on target (orange curve). (c) Demagnification $D$ for different ratios between the XUV source radius and the NIR beam waist radius, showing that the assumption of the XUV source radius has little influence on the anticipated behavior for short NIR focal lengths.}
	\label{raytracing}
\end{figure}

In the calculations, we have taken into account that our laboratory allows a maximum distance of 18~m between the last NIR mirror preceding the HHG source and the mirror that focuses the XUV radiation onto the target. Furthermore, an XUV mirror with a focal length of $f_{XUV}=75$~mm was assumed, which is the shortest focal length that can be used in combination with our velocity-map imaging spectrometer (VMIS). In addition, we have assumed that the XUV source radius is 50~$\%$ of the NIR beam waist radius.

The results of the calculations are shown in Fig.~\ref{raytracing}. The XUV beam waist radius in the experiment (blue curve in Fig.~\ref{raytracing}(a)) increases rapidly with increasing NIR focal length $d_{NIR}$, both due to the increasing XUV source size (as a result of the increasing NIR focus) and the decreasing demagnification. The demagnification calculated by Gaussian optics (solid orange curve) is to a good approximation equal to $D\approx d/f_{XUV}$ for short focal lengths (dotted orange curve), where $d$ is the distance from the HHG source to the XUV mirror. The two curves based on the different approaches deviate for long NIR focal lengths where the XUV beam has a very low divergence. This can be understood by considering the limit of an almost collimated XUV beam, when using an extremely loosely focused NIR beam: In this case, the demagnification would increase with $d_{NIR}$ because of the increasing XUV source size. In our experimental setup, we chose $d\approx 13$~m. This results in a demagnification of 173, independent of whether we assume Gaussian optics or demagnification according to simple geometrical optics.

The global HHG scaling laws presented in Ref.~\cite{heyl16} show that a constant HHG conversion efficiency can be obtained following the change of one parameter (e.g. the NIR pulse energy), if all other relevant parameters (including the NIR focal length and the NIR beam size before focusing) are changed accordingly. In particular, the NIR intensity needs to be kept constant both before focusing (to avoid damage of the focusing optics) and at the focus (for optimal phase matching). This results in an expected linear increase of the XUV pulse energy as a function of the NIR focal length~\cite{heyl16}, as shown in Fig.~\ref{raytracing}(b) (blue curve). The expected XUV intensity on target (orange curve) shows opposite behavior: It increases with decreasing $f_{NIR}$, since the smaller XUV focal spot size more than compensates for the lower XUV energy that is obtained. In Fig.~\ref{raytracing}(c) we show how the demagnification depends on the size of the ratio between the XUV source radius that is assumed and the NIR beam waist radius. Importantly, the demagnification for long NIR focal lengths changes quite drastically with the ratio of the XUV and NIR beam waists, while it is hardly affected for short NIR focal lengths. As a consequence, our results are valid for a broad range of XUV source sizes, a quantity that is not always easy to determine experimentally.

According to these scaling laws, the use of a very short NIR focal length can be used to optimize the focused XUV intensity. However, under such conditions the divergence of the XUV beam is substantial, and restrictions due to finite XUV filter and mirror apertures (which cannot be made arbitrarily large) have to be taken into account, while at the same time the requirements on the quality of the XUV focusing optics increase. As a compromise, we have performed first experiments using $d\approx 13$~m. Besides offering a large demagnification, a further advantage of our beamline geometry is the occurrence of a comparably low NIR intensity on the Al filter that is used to block the NIR light co-propagating with the XUV beam. This removes the need for additional mirrors~\cite{takahashi04} which are often used in high-power HHG applications to attenuate the NIR intensity.

\section{Experimental setup}

\begin{figure*}[!htb]
	\centering
	\fbox{\includegraphics[width=\linewidth]{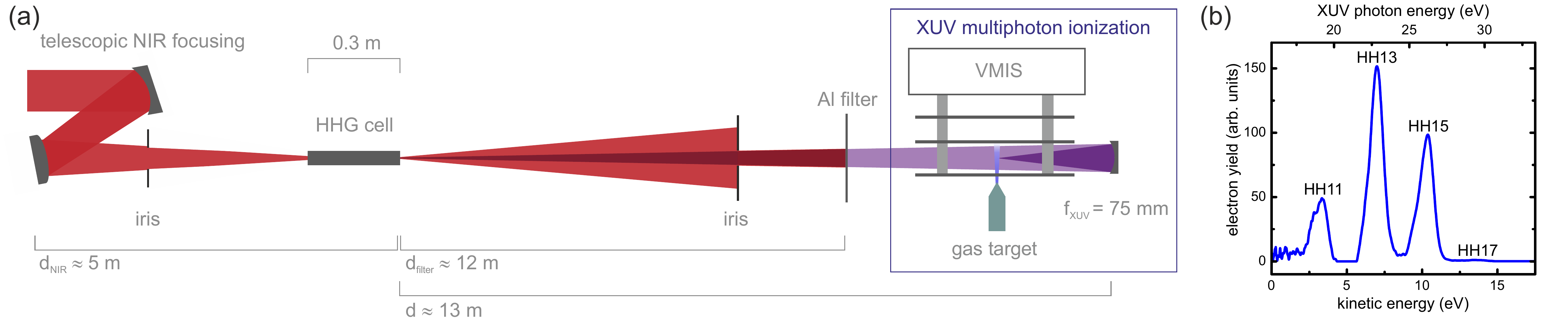}}
	\caption{(a) Experimental setup, see text for details. (b) Photoelectron spectrum of Ar produced by the XUV pulses in the velocity-map imaging spectrometer, showing contributions due to the ionization of high harmonics (HH) of neutral Ar with orders between 11 and 17.}
	\label{setup}
\end{figure*}

The experiments were performed at a recently installed HHG beamline at the Max-Born-Institut. The driving laser concept is based on optical parametric chirped pulse amplification (OPCPA) of a broadband seed beam. The pulses originate from a commercial frontend~\cite{budriunas15} and have a spectrum ranging from 675~nm to 1025~nm. Three high-energy pump beams centered at 515~nm with a total energy $>180$~mJ are generated via frequency doubling of the output from two thin-disk regenerative amplifiers. Amplification is achieved in three stages utilizing beta barium borate (BBO) as a parametric gain medium to yield signal pulse energies up to 42~mJ at a repetition rate of 100~Hz~\cite{kretschmar19}. The initially negatively chirped pulses are compressed using positive material dispersion followed by positively chirped mirrors that are placed in a vacuum chamber. The spectral phase can further be adjusted using an acousto-optic programmable dispersive filter (DAZZLER) placed in the seed beam, which enables a temporal compression to a pulse duration $<9$~fs. The NIR driving pulse energy used for HHG can be varied by changing the pump energy.

Our experimental setup is shown in Fig.~\ref{setup}. We chose a telescopic focusing setup for the NIR pulses using a concave mirror ($f=1.5$~m) and a convex mirror ($f=-1$~m), which has three advantages compared to using a single NIR focusing mirror: (i) Due to the decreased beam size on the last NIR mirror, a large f-number of about 400 can be achieved, along with a comparably large NIR beam waist radius of 270~$\mu$m, thereby increasing the high-harmonic generation volume. (ii) The distance between the last mirror and the NIR focal plane can be easily adjusted by varying the distance between the concave and the convex mirror. (iii) Astigmatism can be compensated by optimizing the reflection angle on the telescope mirrors.

In the setup, the distance between the last NIR mirror and the NIR focal plane is about 5~m. A motorized iris is placed behind the last NIR mirror to optimize the HHG yield, and an NIR driving energy of 11~mJ is measured behind this iris (for an aperture of 10~mm). A 30~cm long gas cell is placed 10~cm in front of the focus, and is statically filled with Xe using a backing pressure of 7~mbar. Gas dynamics simulations were performed, showing that the gas pressure within the cell is about an order of magnitude lower. A comparison with HHG calculations~\cite{major20} indicates a gas pressure in the cell of about 0.5~mbar.

A 100~nm thick Al filter is placed 12~m downstream from the HHG cell and is used to block the co-propagating NIR light. The XUV pulses are focused into the interaction zone of the VMIS using a spherical B$_4$C-coated mirror with a focal length of 75~mm. Argon atoms are injected into the instrument using a piezoelectric valve. The central part of the atomic beam is selected by a molecular beam skimmer with an orifice diameter of 0.5~mm. Ion distributions are recorded in a spatially-resolved manner along the XUV propagation axis using the spatial imaging mode of the VMIS~\cite{eppink97}. To this end, the multichannel plate (MCP) used in our VMIS is gated, and the spatial distributions of the ionic charge states are recorded according to their specific flight times. In addition, the HHG beam diameter can be measured by moving a planar grating into the beam path and steering the zeroth diffraction order onto an MCP / phosphor screen assembly that is placed 12.5~m downstream from the HHG cell.

\section{Estimation of the XUV intensity}

An XUV pulse energy of 56~nJ was measured behind two 100~nm Al filters using an XUV photodiode (AXUV100G) placed 12.5~m downstream from the HHG cell. This corresponds to an estimated XUV pulse energy of 0.7~$\mu$J directly behind the gas cell. We note that a new Al filter with thinner oxide layers was used for measuring the ion distributions after this characterization measurement, resulting in a higher transmission of 40~$\%$. Taking into account an average reflectivity of the XUV focusing mirror of about 25~$\%$, the XUV pulse energy $E$ in the VMIS experiment is expected to be $\approx 70$~nJ. We have further estimated an XUV beam waist radius $w_0$ as 1.3~$\mu$m (see below). The nonlinearity of the HHG process is expected to lead to the production of XUV pulses that are substantially shorter than the duration of the NIR driving pulse. Indeed, previous measurements have reported HHG pulse durations which are shorter than the fundamental pulse duration by factors ranging from 1.6 to 2.9~\cite{mauritsson04, nabekawa05}. Accordingly, assuming an XUV pulse duration of $\Delta t=4$~fs, the XUV peak intensity is estimated as $I_{peak}=2E/(\Delta t \pi w_0^2)\approx 7\times 10^{14}~$W/cm$^2$.

The beam waist radius of 1.3~$\mu$m was inferred from spatially resolved measurements of the formation of multiply charged Ar ions that will be discussed in Sec.~5. An even smaller XUV beam waist radius is suggested by calculations that were performed to study HHG under the conditions of our experiment~\cite{major20}. The results of these calculations show that reshaping of the driving laser in the HHG medium plays an important role, resulting in efficient, phase-matched HHG. Furthermore, a short attosecond pulse train was predicted in the focus of the experiment~\cite{major20}. Taking into account the attosecond pulse structure and the prediction of an XUV beam waist radius of 320~nm that was found in the calculations, an even higher XUV intensity of $9 \times 10^{15}$~W/cm$^2$ may be reached on target.

\section{Multiphoton ionization of Ar atoms}

\begin{figure}[!htb]
\centering
\fbox{\includegraphics[width=0.75\linewidth]{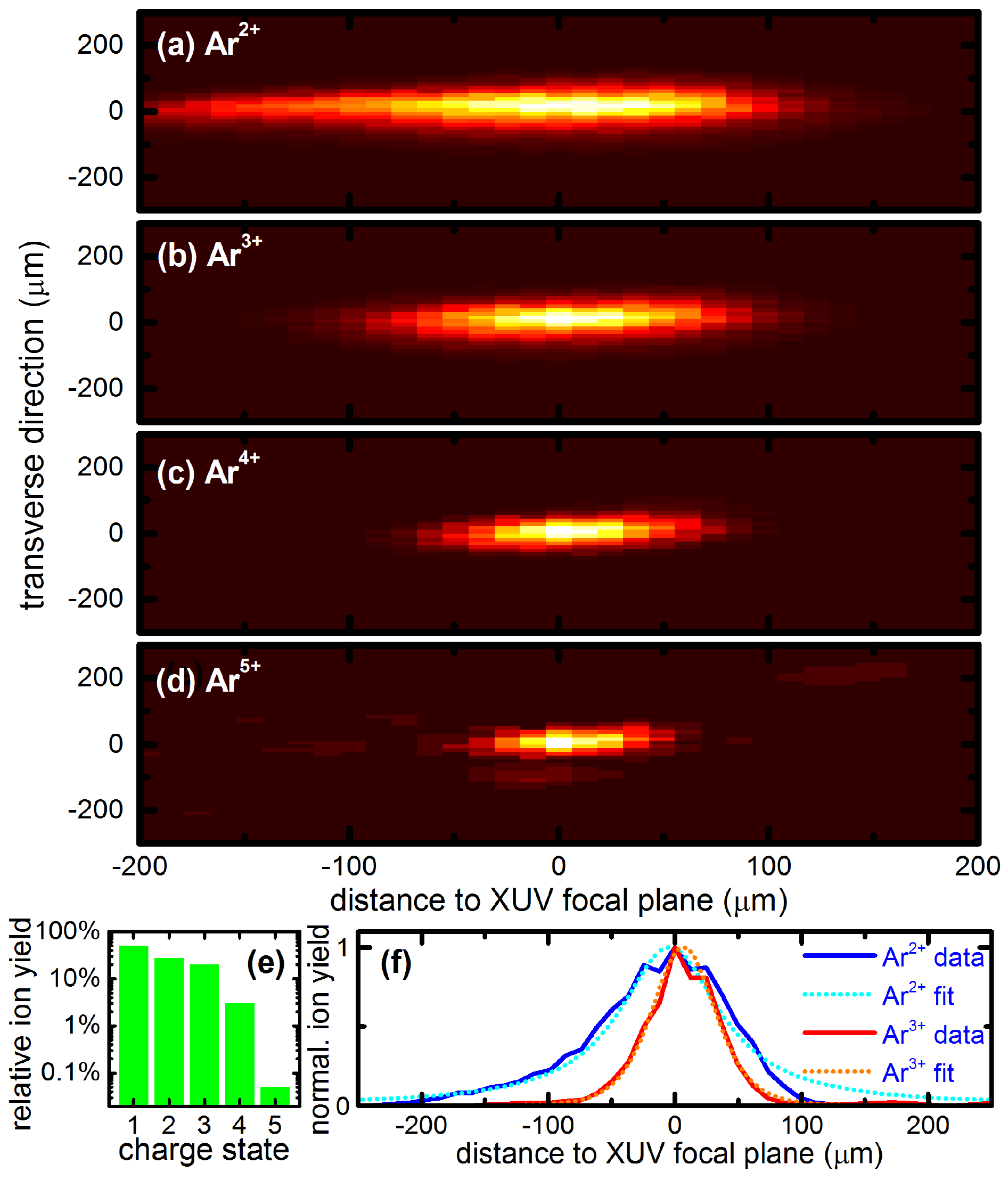}}
\caption{Measured spatial distributions of (a) Ar$^{2+}$, (b) Ar$^{3+}$, (c) Ar$^{4+}$ and (d) Ar$^{5+}$ ions, where the horizontal axis refers to the XUV propagation axis and the vertical axis refers to the transverse direction. Each ion distribution was normalized to its maximal value. Note that the extension in the transverse direction is limited by the spatial resolution of the imaging setup of about $50~\mu$m. Note further that some noise is visible in (d) due to the low signal level. (e) Relative ion contributions at the XUV focal plane. (f) The solid curves represent transversely integrated normalized Ar$^{2+}$ and Ar$^{3+}$ ion distributions. In these measurements a second Al filter was inserted to avoid saturation of the individual ion yields. The dotted curves are fits to the ion distributions to extract the Rayleigh lengths, see text for details.}
\label{multiphoton-ionization}
\end{figure}

To demonstrate and investigate the utility of the focused XUV pulses for XUV nonlinear optics, CDI and XUV pump - XUV probe experiments, we have studied multiphoton ionization of Ar atoms, and observed the generation of ions with charge states up to Ar$^{5+}$. The spatial ion distributions along the XUV propagation direction (Fig.~\ref{multiphoton-ionization}) are peaked at the XUV focal plane and become narrower for more highly charged ions, which is a signature of the higher orders of nonlinearity leading to the production of these ions. We note that the absence of sidebands in the electron spectrum shown in Fig.~2(b) demonstrates that the residual NIR light did not play a role in the ionization of Ar. The spatially resolved measurements allow us to determine the relative contributions of the different ion species in the focal plane (Fig.~\ref{multiphoton-ionization}(e)), showing significant contributions of Ar$^{2+}$ (27.5~$\%$), Ar$^{3+}$ (19.5~$\%$) and Ar$^{4+}$ (3~$\%$) ions as well as a small contribution of Ar$^{5+}$ (0.05~$\%$) ions. These results are comparable to previously obtained FEL results~\cite{motomura09}. 

Since it is not easily possible to measure the XUV beam waist radius when using a spherical focusing mirror with a short focal length, we have instead made an estimation based on the XUV beam size on the focusing mirror and an estimate of the Rayleigh length $z_R$ making use of the longitudinal Ar$^{2+}$ and Ar$^{3+}$ ion distributions (solid curves in Fig.~\ref{multiphoton-ionization}(f)). For this measurement, we have inserted a second Al filter with a transmission of 16~$\%$ to avoid significant depletion of the Ar$^{2+}$ and Ar$^{3+}$ ion yields due to the formation of more highly charged ions. The transversely integrated ion yield scales with $I^n(z) \times w^2(z) \propto I^n/I= I^{n-1}$, where $I(z)\propto 1/w^2(z)$ is the XUV intensity at a distance $z$ from the focal plane, $n$ is the order of nonlinearity, and $w(z)=w_0\sqrt{1+z^2/z_R^2}$ is the beam radius, with $w_0$ being the beam waist radius. Accordingly, the ion distributions are proportional to $(1+z^2/z_R^2)^{-(n-1)}$. As indicated by measurements of the intensity dependence of the ion yields (see below), the generation of Ar$^{2+}$ is described by an effective nonlinearity of $n=2$, while the generation of Ar$^{3+}$ is described by an effective nonlinearity of $n=4$. Fits using these orders of nonlinearity are shown as dotted curves in Fig.~\ref{multiphoton-ionization}(f), giving $z_R=49$~$\mu$m for the fit of the Ar$^{2+}$ and $z_R=58$~$\mu$m for the fit of the Ar$^{3+}$ result. Therefore, we estimate that $z_R=54\pm 8$~$\mu$m. The XUV beam radius on the MCP / phosphor screen was measured to be 1.7~mm, indicating an XUV beam radius $w_{XUV,mirror}=1.8$~mm at the position of the XUV focusing mirror. The beam waist radius is estimated according to $w_0=w_{XUV,mirror} (1+f_{XUV}^2/z_R^2)^{-1/2}\approx w_{XUV,mirror} z_R/f_{XUV}$, giving a value of $1.3\pm 0.2$~$\mu$m. Again, we point out that numerical calculations using the parameters of our beamline~\cite{major20} indicate a significantly smaller XUV beam waist radius of 320~nm as a result of reshaping of the driving pulses in the HHG cell. In our experiment, we would not be able to resolve such a small focus size, which is associated with a Rayleigh of only several micrometers~\cite{major20}, due to the limited spatial resolution of our setup of about 50~$\mu$m. We further note that the asymmetric behavior of the Ar$^{2+}$ ion yield before and after the focus observed in Fig.~\ref{multiphoton-ionization}(a) may be a result of propagation effects of the driving laser in the HHG cell, which are described in detail in Ref.~\cite{major20}.

\begin{figure}[tb]
\centering
\fbox{\includegraphics[width=0.5\linewidth]{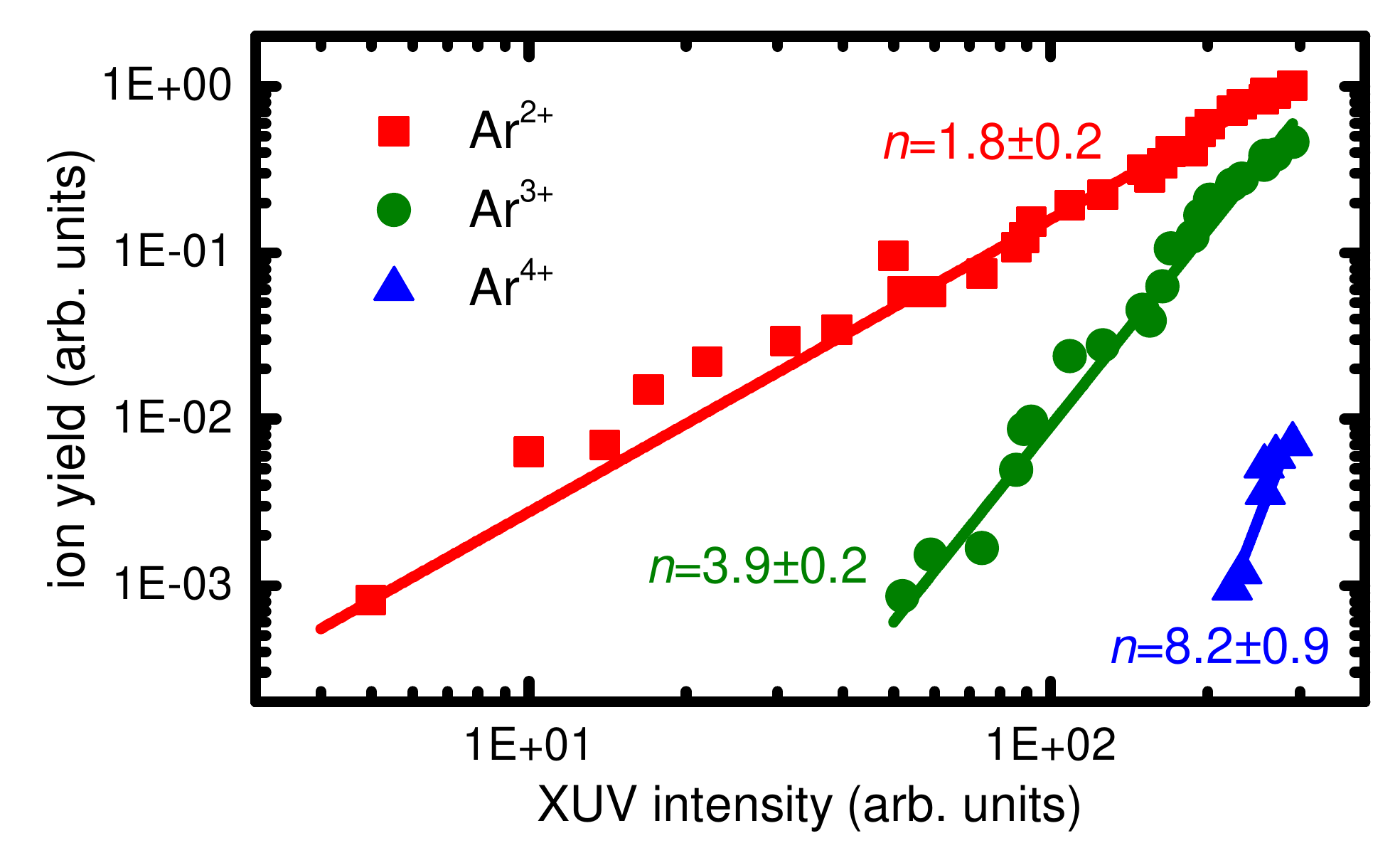}}
\caption{Ion yields as a function of the XUV intensity (symbols) that was varied by changing the gas pressure in the HHG cell. The Ar$^+$ ion yield observed far away from the focus was used to calibrate the relative XUV intensities. Fits to the different ion species (solid curves) indicate orders of nonlinearities $n$ of $1.8\pm 0.2$ for Ar$^{2+}$, $3.9\pm 0.2$ for Ar$^{3+}$ and $8.2\pm 0.9$ for Ar$^{4+}$. Note that the maximal XUV peak intensity in this measurement was somewhat lower compared to the measurement shown in Fig.~3.}
\label{intensity-dependence}
\end{figure}

Fig.~4 shows the intensity-dependent Ar$^{2+}$, Ar$^{3+}$ and Ar$^{4+}$ ion yields, from which we can extract the orders of nonlinearity of the different ion species. We find $n=1.8\pm 0.2$ for Ar$^{2+}$, $n=3.9 \pm 0.2$ for Ar$^{3+}$ and $n=8.2 \pm 0.9$ for Ar$^{4+}$, indicating that about 8 XUV photons are absorbed in the latter case. While we have not recorded intensity-dependent data for Ar$^{5+}$ due to the low signal, we estimate that 3-4 additional photons are required to generate Ar$^{5+}$ from the Ar$^{4+}$ ground state, which has an ionization potential of 74.8~eV~\cite{biemont99}. From this consideration it follows that at least 10 HHG photons are absorbed by a single photon.

In the production of these ions, both sequential and nonsequential absorption of XUV photons may contribute. Since our photon energy range is below the ionization potentials of more highly charged ions, absorption of XUV photons leading to the formation of excited ions is expected to play an important role. The subsequent absorption of an additional XUV photon may then lead to further ionization of these excited ions. Nonsequential ionization following the simultaneous absorption of two or more XUV photons is expected to become important at very short XUV pulse durations~\cite{tzallas11}. Future XUV-XUV experiments will allow us to identify the roles that are played by sequential and nonsequential ionization, since the latter will result in an increased ion yield at the temporal overlap of the two XUV pulses.

We note that the production of ions up to Ar$^{4+}$ was observed in a recent experiment using HHG, in which XUV pulses with energies of up to 230~$\mu$J were reported~\cite{nayak18}. Our findings demonstrate that we achieve a higher degree of ionization, although our XUV pulse energy is more than two orders of magnitude lower than the pulse energy reported in Ref.~\cite{nayak18}. Our results therefore open a path for the investigation of highly nonlinear processes in the XUV range using HHG sources that are driven by lasers with moderately high pulse energies, which makes it feasible to perform these experiments using lasers at kHz repetition rates in the future.

\section{Summary}

In summary, we have demonstrated the generation of XUV pulses by HHG reaching intensities of at least $7\times 10^{14}$~W/cm$^2$ on target using a novel scaling scheme. These XUV pulses enable highly nonlinear ionization of Ar atoms up to Ar$^{5+}$, following the absorption of at least 10 XUV photons. Our results provide excellent opportunities for XUV pump - XUV probe experiments and for the single-shot CDI of isolated nanotargets.

\section*{Acknowledgments}

Funding by the Leibniz Grant No. SAW/2017/MBI4 is ackowledged. We are grateful for the technical support by M. Krause, C. Reiter, W. Krüger and R. Peslin.

\section*{References}
% Bibliography

%\bibliography{References}

\begin{thebibliography}{10}
\expandafter\ifx\csname url\endcsname\relax
  \def\url#1{{\tt #1}}\fi
\expandafter\ifx\csname urlprefix\endcsname\relax\def\urlprefix{URL }\fi
\providecommand{\eprint}[2][]{\url{#2}}

\bibitem{jiang10}
Jiang Y~H, Rudenko A, Herrwerth O, Foucar L, Kurka M, K\"uhnel K~U, Lezius M,
  Kling M~F, van Tilborg J, Belkacem A, Ueda K, D\"usterer S, Treusch R,
  Schr\"oter C~D, Moshammer R and Ullrich J 2010 {\em Phys. Rev. Lett.\/} {\bf
  105}(26) 263002
  \urlprefix\url{https://link.aps.org/doi/10.1103/PhysRevLett.105.263002}

\bibitem{tzallas11}
Tzallas P, Skantzakis E, Nikolopoulos L, Tsakiris G~D and Charalambidis D 2011
  {\em Nat. Phys.\/} {\bf 7} 781
  \urlprefix\url{https://doi.org/10.1038/nphys2033}

\bibitem{takahashi13}
Takahashi E~J, Lan P, M{\"u}cke O~D, Nabekawa Y and Midorikawa K 2013 {\em Nat.
  Commun.\/} {\bf 4} 2691
  \urlprefix\url{https://www.nature.com/articles/ncomms3691}

\bibitem{bogan08}
Bogan M~J, Benner W~H, Boutet S, Rohner U, Frank M, Barty A, Seibert M~M, Maia
  F, Marchesini S, Bajt S {\em et~al.\/} 2008 {\em Nano Lett.\/} {\bf 8}
  310--316 \urlprefix\url{10.1021/nl072728k}

\bibitem{rupp17}
Rupp D, Monserud N, Langbehn B, Sauppe M, Zimmermann J, Ovcharenko Y,
  M{\"o}ller T, Frassetto F, Poletto L, Trabattoni A {\em et~al.\/} 2017 {\em
  Nat. Commun.\/} {\bf 8} 493
  \urlprefix\url{https://www.nature.com/articles/s41467-017-00287-z}

\bibitem{motomura09}
Motomura K, Fukuzawa H, Foucar L, Liu X~J, Pr\"umper G, Ueda K, Saito N,
  Iwayama H, Nagaya K, Murakami H, Yao M, Belkacem A, Nagasono M, Higashiya A,
  Yabashi M, Ishikawa T, Ohashi H and Kimura H 2009 {\em J. Phys. B\/} {\bf 42}
  221003 \urlprefix\url{https://doi.org/10.1088%2F0953-4075%2F42%2F22%2F221003}

\bibitem{foglia18}
Foglia L, Capotondi F, Mincigrucci R, Naumenko D, Pedersoli E, Simoncig A,
  Kurdi G, Calvi A, Manfredda M, Raimondi L {\em et~al.\/} 2018 {\em Phys. Rev.
  Lett.\/} {\bf 120}(26) 263901
  \urlprefix\url{https://link.aps.org/doi/10.1103/PhysRevLett.120.263901}

\bibitem{savelyev17}
Savelyev E, Boll R, Bomme C, Schirmel N, Redlin H, Erk B, D{\"u}sterer S,
  M{\"u}ller E, H{\"o}ppner H, Toleikis S {\em et~al.\/} 2017 {\em New J.
  Phys.\/} {\bf 19} 043009
  \urlprefix\url{https://doi.org/10.1088%2F1367-2630%2Faa652d}

\bibitem{krausz09}
Krausz F and Ivanov M 2009 {\em Rev. Mod. Phys.\/} {\bf 81}(1) 163--234
  \urlprefix\url{https://link.aps.org/doi/10.1103/RevModPhys.81.163}

\bibitem{schutte14a}
Sch{\"u}tte B, Arbeiter M, Fennel T, Vrakking M~J and Rouz{\'e}e A 2014 {\em
  Phys. Rev. Lett.\/} {\bf 112}(7) 073003
  \urlprefix\url{https://link.aps.org/doi/10.1103/PhysRevLett.112.073003}

\bibitem{manschwetus16}
Manschwetus B, Rading L, Campi F, Maclot S, Coudert-Alteirac H, Lahl J, Wikmark
  H, Rudawski P, Heyl C, Farkas B {\em et~al.\/} 2016 {\em Phys. Rev. A\/} {\bf
  93}(6) 061402
  \urlprefix\url{https://link.aps.org/doi/10.1103/PhysRevA.93.061402}

\bibitem{nayak18}
Nayak A, Orfanos I, Makos I, Dumergue M, K{\"u}hn S, Skantzakis E, Bodi B,
  Varju K, Kalpouzos C, Banks H {\em et~al.\/} 2018 {\em Phys. Rev. A\/} {\bf
  98}(2) 023426
  \urlprefix\url{https://link.aps.org/doi/10.1103/PhysRevA.98.023426}

\bibitem{bergues18}
Bergues B, Rivas D~E, Weidman M, Muschet A~A, Helml W, Guggenmos A, Pervak V,
  Kleineberg U, Marcus G, Kienberger R, Charalambidis D, Tzallas P,
  Schr\"{o}der H, Krausz F and Veisz L 2018 {\em Optica\/} {\bf 5} 237--242
  \urlprefix\url{http://www.osapublishing.org/optica/abstract.cfm?URI=optica-5-3-237}

\bibitem{boutu11}
Boutu W, Auguste T, Caumes J~P, Merdji H and Carr\'e B 2011 {\em Phys. Rev.
  A\/} {\bf 84}(5) 053819
  \urlprefix\url{https://link.aps.org/doi/10.1103/PhysRevA.84.053819}

\bibitem{heyl16}
Heyl C~M, Arnold C~L, Couairon A and L'Huillier A 2016 {\em J. Phys. B\/} {\bf
  50} 013001
  \urlprefix\url{https://doi.org/10.1088%2F1361-6455%2F50%2F1%2F013001}

\bibitem{siegman86}
Siegman A 1986 {\em Lasers\/} (University Science Books) ISBN 9780935702118
  \urlprefix\url{https://books.google.de/books?id=1BZVwUZLTkAC}

\bibitem{porras92}
Porras M~A, Alda J and Bernabeu E 1992 {\em Appl. Opt.\/} {\bf 31} 6389--6402
  \urlprefix\url{http://ao.osa.org/abstract.cfm?URI=ao-31-30-6389}

\bibitem{takahashi04}
Takahashi E~J, Hasegawa H, Nabekawa Y and Midorikawa K 2004 {\em Opt. Lett.\/}
  {\bf 29} 507--509
  \urlprefix\url{http://ol.osa.org/abstract.cfm?URI=ol-29-5-507}

\bibitem{budriunas15}
Budri{\={u}}nas R, Stanislauskas T and Varanavi{\v{c}}ius A 2015 {\em J.
  Opt.\/} {\bf 17} 094008
  \urlprefix\url{https://doi.org/10.1088%2F2040-8978%2F17%2F9%2F094008}

\bibitem{kretschmar19}
Kretschmar M {\em et~al.\/} {\em in preparation\/}

\bibitem{major20}
Major B, Kretschmar M, Ghafur O, Hoffmann A, Kovács K, Varjú K, Senfftleben
  B, Tümmler J, Will I, Nagy T, Rupp D, Vrakking M~J~J, Tosa V and Schütte B
  2020 Propagation-assisted generation of intense few-femtosecond high-harmonic
  pulses (\textit{Preprint} \eprint{2002.07139})
  \urlprefix\url{https://arxiv.org/abs/2002.07139}

\bibitem{eppink97}
Eppink A~T~J~B and Parker D~H 1997 {\em Rev. Sci. Instrum.\/} {\bf 68} 3477
  \urlprefix\url{https://doi.org/10.1063/1.1148310}

\bibitem{mauritsson04}
Mauritsson J, Johnsson P, L{\'o}pez-Martens R, Varju K, Kornelis W, Biegert J,
  Keller U, Gaarde M, Schafer K and L'Huillier A 2004 {\em Phys. Rev. A\/} {\bf
  70}(2) 021801
  \urlprefix\url{https://link.aps.org/doi/10.1103/PhysRevA.70.021801}

\bibitem{nabekawa05}
Nabekawa Y, Hasegawa H, Takahashi E~J and Midorikawa K 2005 {\em Phys. Rev.
  Lett.\/} {\bf 94}(4) 043001
  \urlprefix\url{https://link.aps.org/doi/10.1103/PhysRevLett.94.043001}

\bibitem{biemont99}
Bi{\'E}mont E, Fr{\'e}mat Y and Quinet P 1999 {\em At. Data and Nucl. Data
  Tables\/} {\bf 71} 117--146
  \urlprefix\url{https://doi.org/10.1006/adnd.1998.0803}

\end{thebibliography}

\providecommand{\newblock}{}

\end{document}